\documentclass[a4paper,fleqn,usenatbib]{mnras}

\usepackage[T1]{fontenc}
\usepackage{ae,aecompl}

\usepackage{graphicx}	
\usepackage{siunitx, tensor, subfig, tabu}
\usepackage[stable]{footmisc}
\usepackage[inline]{enumitem}
\graphicspath{ {figures//} }

\defcitealias{poci2021}{P21}

\usepackage{cmdDef}

    \title[F3D: Orbital IMF]{The Fornax3D project: Intrinsic Correlations between Orbital Properties and the Stellar Initial Mass Function}

    \author[A. Poci et al.]{%
        A.~Poci$^{1, 2}$\thanks{E-mail: adriano.poci@durham.ac.uk (DU)},
        R.~M.~McDermid$^{2, 3}$,
        M.~Lyubenova$^{4}$,
        I.~Mart{\'i}n-Navarro$^{5, 6}$,
        G.~van de Ven$^{7}$,\and
        L.~Coccato$^{4}$,
        E.~M.~Corsini$^{8, 9}$,
        K.~Fahrion$^{10}$,
        J.~Falc{\'o}n-Barroso$^{5, 6}$,
        D.~A.~Gadotti$^{4}$,\and
        E.~Iodice$^{11}$,
        F.~Pinna$^{12}$,
        M.~Sarzi$^{13, 14}$,
        P.~T.~de~Zeeuw$^{15, 16}$,
        L.~Zhu$^{17}$\\
        $^{1}$Centre for Extragalactic Astronomy, University of Durham, Stockton Road, Durham DH1 3LE, United Kingdom\\
        $^{2}$Research Centre for Astronomy, Astrophysics, and Astrophotonics, Department of Physics and Astronomy, Macquarie University,\\NSW 2109, Australia\\
        $^{3}$ARC Centre of Excellence for All Sky Astrophysics in 3 Dimensions (ASTRO 3D), Australia\\
        $^{4}$European Southern Observatory, Karl-Schwarzschild-Stra{\ss}e 2, D-85748 Garching bei M{\"u}nchen, Germany\\
        $^{5}$Instituto de Astrof{\'i}sica de Canarias, Calle Via L{\'a}ctea s/n, E38205 La Laguna, Tenerife, Spain\\
        $^{6}$Departamento de Astrof{\'i}sica, Universidad de La Laguna, E38205 La Laguna, Tenerife, Spain\\
        $^{7}$Department of Astrophysics, University of Vienna, T{\"u}rkenschanzstra{\ss}e 17, 1180 Wien, Austria\\
        $^{8}$Dipartimento di Fisica e Astronomia `G. Galilei', Universit{\`a} di Padova, vicolo dell'Osservatorio 3, I-35122 Padova, Italy\\
        $^{9}$INAF - Osservatorio Astronomico di Padova, vicolo dell'Osservatorio 5, I-35122 Padova, Italy\\
        $^{10}$European Space Agency, European Space Research and Technology Centre, Keplerlaan 1, 2200 AG Noordwijk, Netherlands\\
        $^{11}$INAF–Osservatorio Astronomico di Capodimonte, via Moiariello 16, I-80131 Napoli, Italy\\
        $^{12}$Max-Planck-Institut f{\"u}r Astronomie, K{\"o}nigstuhl 17, 69117 Heidelberg, Germany\\
        $^{13}$Armagh Observatory and Planetarium, College Hill, Armagh BT61 9DG, UK\\
        $^{14}$Centre for Astrophysics Research, University of Hertfordshire, College Lane, Hatfield AL10 9AB, UK\\
        $^{15}$Sterrewacht Leiden, Leiden University, Postbus 9513, 2300 RA Leiden, The Netherlands\\
        $^{16}$Max-Planck-Institut f{\"u}r Extraterrestrische Physik, Gie{\ss}enbachstra{\ss}e 1, 85748 Garching bei M{\"u}nchen, Germany\\
        $^{17}$Shanghai Astronomical Observatory, Chinese Academy of Sciences, 80 Nandan Road, Shanghai 200030, China
    }

    \date{Accepted XXX. Received YYY; in original form ZZZ}
    \pubyear{2021}

\begin{document}
\label{firstpage}
\pagerange{\pageref{firstpage}--\pageref{lastpage}}
\maketitle

    \begin{abstract}
    Variations of the stellar Initial Mass Function (IMF) in external galaxies have been inferred from a variety of independent probes. These variations occur not just between but also within galaxies, as evidenced by recent spectral and dynamical analyses. Yet the physical conditions relating to variations in the IMF remaining largely unknown, driven by the difficulty in observationally measuring said variations. In this work, we explore new spatially-resolved measurements of the IMF for three edge-on lenticular galaxies in the Fornax cluster. Specifically, we utilise existing orbit-based dynamical models, which re-produce the measured stellar kinematics, in order to fit the new IMF maps within this orbital framework. We then investigate correlations between intrinsic orbital properties and the local IMF. We find that, within each galaxy, the high-angular-momentum, disk-like stars exhibit an IMF which is rich in dwarf stars. The centrally-concentrated pressure-supported orbits have IMF which are similarly rich in dwarf stars. Conversely, orbits at large radius which have intermediate angular momentum exhibit IMF which are markedly less dwarf-rich relative to the other regions of the same galaxy. Assuming that the stars which, in the present-day, reside on dynamically-hot orbits at large radii are dominated by accreted populations, we can interpret these findings as a correlation between the dwarf-richness of a population of stars, and the mass of the host in which it formed. Specifically, deeper gravitational potentials would produce more dwarf-rich populations, resulting in the relative deficiency of dwarf stars which originated in the lower-mass accreted satellites. Conversely, the central and high angular-momentum populations are likely dominated by {\em in-situ} stars, which were formed in the more massive host itself. There are also global differences between the three galaxies studied here, of up to \(\sim 0.3\ \si{dex}\) in the IMF parameter \(\xi\). We find no local dynamical or chemical property which alone can fully account for the IMF variations.
    \end{abstract}



\section{Introduction}\label{sec:imfIntro}
The stellar Initial Mass Function (IMF) is of paramount importance to many fields of astrophysics. Originally conceived as the probability distribution function of a local star-formation episode as a function of stellar mass \citep[though see][for alternative interpretations]{kroupa2019}, the IMF has a fundamental role across all scales of astrophysics. From star formation in individual Giant Molecular Clouds (GMC), to galaxy formation, and subsequently the assembly of structure in the Universe on large scales, the (potentially redshift-dependent) IMF has an impact across a broad range of physical and temporal scales. The IMF has direct implications for chemical yields from stellar evolution  and chemical evolution on galactic scales \citep[e.g.][]{yan2020}, so an accurate IMF is critical to test models of those processes.\par
Despite its importance for extragalactic measurements, the vast majority of work on external galaxies is under the assumption of a constant IMF; between galaxies, within galaxies, or both. This is primarily driven by technical and fundamental limitations associated with {\em measuring} the IMF -- especially for external galaxies. Yet indirect methods have provided mounting evidence of variations of the IMF between individual galaxies, and typically with some dependence on integrated galactic properties such as the velocity dispersion \citep{treu2010, thomas2011, cappellari2012, labarbera2013, spiniello2014, tortora2014, posacki2015, li2017, rosani2018}, \(\alpha\)-element enrichment \citep{conroy2012, mcdermid2014}, or total metal enrichment \citep{martin-navarro2015}.\par
With increasingly modern data and measurement techniques, it is also becoming increasingly clear that variations of the IMF are local and correlate with local galactic properties. This is beginning to be exploited already by improved analysis techniques. Using spatially-resolved data, gravity-sensitive spectral features --- which provide constraints on the relative abundance of low-mass stars, which in turn is related to the slope of the IMF --- are observed to vary with galacto-centric radius, becoming relatively dwarf-rich at low radius \citep{martin-navarro2015a, martin-navarro2015b, labarbera2016, vandokkum2017}. In a similar analysis, \cite{parikh2018} find similar radial trends --- increasing dwarf-richness with decreasing radius ---  and an additional stellar mass dependence of that radial trend. They find, however, that using only the Wing-Ford \(\chemFH\) gravity-sensitive absorption feature produces the inverse radial behaviour, highlighting the observational difficulties associated with constraining the IMF. \cite{sarzi2018} also found increasing dwarf-richness with decreasing radius with an abundance analysis of the nearby massive galaxy M 87 (NGC 4486). \cite{labarbera2019} find that radius and surface mass density exhibit the strongest correlations with gravity-sensitive spectral features, even compared to the velocity dispersion. \par
Providing orthogonal constraints to gravity-sensitive absorption feature analyses, the discrepancy between dynamical and stellar masses (for a given IMF), the so-called IMF mis-match parameter \(\alpha_{\rm IMF}\), is interpreted as a measure of the IMF, provided the dark matter can be accurately taken into account. Variations in this parameter are seen as evidence for a non-universal IMF both between and within galaxies \citep[e.g.][]{cappellari2012, lyubenova2016, li2017, oldham2018}. Using molecular gas as a high-spatial-resolution tracer of the gravitational potential (dynamical mass) in the central regions of a sample of nearby galaxies, \cite{davis2017} find that this discrepancy can correlate positively, negatively, or not at all with galacto-centric radius. They also report no correlations with other galactic properties, global or local.\par
The emphasis in recent times has been on the development of statistically-robust methods, increasingly exploiting Bayesian techniques. This was motivated by the large number of free parameters of increasingly-complex models, as well as the need to characterise potential correlations between them. In the specific case of the IMF, this approach has had a number of implementations. These include {\tt alf} \citep{conroy2012a, conroy2013, conroy2017}, {\tt PyStaff} \citep{vaughan2018a}, and ``Full-Index Fitting'' \citep[FIF;][]{martin-navarro2019}. This additional statistical constraining power comes with a marked increase in computational cost, so most of these techniques are restricted to coarse radial bins rather than taking advantage of the truly spatially-resolved IFU data (though see below).\par
While this transition from global to local correlations of the IMF represents a dramatic improvement in extragalactic IMF studies, the physical cause of these correlations remains elusive. Understanding what drives radial variations of the IMF has been challenging due to the vast array of galactic properties which also vary with radius, in particular in the massive ETG which are most often the focus of extragalactic IMF studies. Unlike its global counterpart, the local velocity dispersion has been shown to be a poor predictor of the local IMF \citep[e.g.][]{martin-navarro2015, barbosa2021a}. Metallicity is key property believed to impact the IMF, and it often exhibits qualitatively similar (radially declining) gradients as observed for the IMF. However, the direct link between metallicity and IMF is not so clear; the correlation exhibits significant scatter and may be different for individual galaxies \citep{martin-navarro2021a}. Moreover, metallicity variations alone can not account for the observed IMF variations \citep[e.g.][]{martin-navarro2015, martin-navarro2021a, mcconnell2016, villaume2017a}. As such, it remains unclear with which galactic properties the IMF fundamentally varies.\par
Evidently these measurements require detailed modelling across the parameter-space of galactic properties in order to uncover drivers of this variation. In this work, we build upon the results of the \ftd\ survey, which has presented spatially-resolved maps of the stellar IMF for quiescent galaxies in the survey \citep{martin-navarro2021a}, as well as a qualitative connection between the structures of the projected IMF maps and the projected orbital components \citep{martin-navarro2019}. In the present work, we apply a powerful orbit-based population-dynamical model to the FIF method for measuring the variations of the IMF across individual galaxies. By re-producing the measured IMF maps using our orbital dynamical model, we aim to investigate quantitative correlations of the local intrinsic kinematics with the local IMF.

\section{Data and models}\label{sec:data}
In this work, we make direct use of the data and analysis presented in \citet[][hereafter \citetalias{poci2021}]{poci2021}. That work studied three lenticular galaxies -- FCC~153, FCC~170, and FCC~177 -- using spectroscopic data from the \ftd\ survey \citep{sarzi2018} and photometric data from the Fornax Deep Survey \citep[FDS;][]{iodice2016, venhola2018}. These galaxies are ideal because their nearly edge-on projection minimises any uncertainty in analysing the various galactic components. Stellar kinematics were measured by extracting the first \(6\) Gauss-Hermite coefficients of the line-of-sight velocity distribution (LOSVD), fitting the observed spectra with the {\rm MILES} empirical stellar library \citep{sanchez-blazquez2006, falcon-barroso2011} in the \tfo{pPXF} \citep{cappellari2004, cappellari2017} Python package\footnote{Available at \href{https://pypi.org/project/ppxf/}{https://pypi.org/project/ppxf/}}. Stellar populations (ages, metallicities, and stellar mass-to-light ratios) were measured by fitting the observed spectra with the {\rm E-MILES} single stellar population (SSP) library \citep{vazdekis2016}, varying in age and metallicity for a fixed \cite{kroupa2002} IMF, again using \tfo{pPXF}. The analysis of these data comprised of \shw\ orbit-based dynamical models \citep[using a general triaxial implementation;][]{vandeven2008, vandenbosch2008}\footnote{we confirmed that the issues reported in \protect\cite{quenneville2022} do not affect our best-fit models. We re-ran the model for FCC~170 and confirmed that none of the best-fit model parameters depart from the values presented in \citetalias{poci2021}. A more thorough comparison is presented in \protect\citep{thater2022}, showing that none of the physical properties of the models are significantly impacted, over a broad range of galaxy types and observational data.}, and the self-consistent fitting of the measured stellar populations with these dynamical models.\par
Specifically, the best-fitting dynamical model for each galaxy was divided into a number of sub-components defined by their intrinsic kinematics properties. These properties are the orbital circularity, \(\lambda_z\) \citep{zhu2018b}, and the cylindrical time-averaged radius \(R\). Each resulting sub-component has known intrinsic kinematic properties, and a known contribution to the original dynamical model. The projected mean stellar age and metallicity measured in \citetalias{poci2021} were then fit using this basis set of orbital components. This produced a model describing all measured stellar properties (kinematics, ages, and metallicities) within a single self-consistent orbital framework. We refer the reader to \citetalias{poci2021} for a full presentation of the dynamical model fits and related details. Here we present the extension of this modelling approach to include an additional stellar-population property; namely, spatially-resolved measurements of the stellar IMF.

\section{Modelling the Stellar Populations}\label{sec:methods}
\subsection{Full-Index Fitting}\label{ssec:imfFIF}
Exploiting the parallel advances in computing technologies and observational instrumentation, \cite{martin-navarro2019} presented measurements which are sensitive to the relative fraction of low-mass stars. This is achieved by targeting specific absorption features that maximise the sensitivity to key parameters (including elemental abundances) while minimising the computational overhead in fitting spectral pixel data. It is sufficiently computationally-efficient that this measurement can be made for each spectrum in an IFU data-cube, and the IMF can therefore be investigated in a truly spatially-resolved manner (not just radially).\par
The IMF in this instance is assumed to be a power-law in stellar mass for high masses, and have a gradient of zero at low masses, following the functional form originally proposed by \cite{vazdekis1996}, and given here in \cref{eq:imf}.
\begin{flalign}\label{eq:imf}
    && \Phi(m) &= \begin{cases}
        \beta\ 0.4^{-\Gamma_b}, & m/\Msun \leq 0.2\\
        \beta\ p(m), & 0.2 < m/\Msun < 0.6\\
        \beta\ m^{-\Gamma_b}, & m/\Msun \geq 0.6
    \end{cases} &&
    \intertext{where \(p(m)\) is a spline ensuring a smooth transition between the two mass regimes, subject to the following boundary constraints:}
    && p(0.2) &= 0.4^{-\Gamma_b} &&\notag\\
    && \frac{\dif p}{\dif m}(0.2) &= 0 &&\notag\\
    && p(0.6) &= 0.6^{-\Gamma_b} &&\notag\\
    && \frac{\dif p}{\dif m}(0.6) &= -\Gamma_b\ 0.6^{-\Gamma_b-1} &&\notag
\end{flalign}
Population synthesis models are generated using this IMF, and the observed spectra are then compared with these models in order to make the measurements. Unlike conventional full-spectral fitting techniques, FIF isolates specific regions of a spectrum which contain IMF-sensitive information. In practise, this amounts to narrow band-passes surrounding key absorption features. FIF treats each spectral pixel along these absorption features as independent data-points. This is therefore similar to full-spectral fitting, but without including the continuum regions between absorption features of the spectra.\par
Specifically, the \(\nuclide{Fe}5270\), \(\nuclide{Fe}5335\), \(\nuclide{Mg\ b}5177\), \(\nuclide{aTiO}\), \(\nuclide{TiO_1}\), and \(\nuclide{TiO_2}\) absorption features are modelled simultaneously to constrain the mean stellar age, metallicity, elemental abundances \(\chemAlphFe\), and the high-mass power-law IMF slope \(\Gamma_b\). The IMF is subsequently re-parametrised in order to emphasise the changes in the stellar populations that are still directly observable in the present day. The specific parametrisation utilised by FIF, and in this work, is defined as
\begin{flalign}
    && \xi &= \frac{\int_{m=0.2}^{0.5} \Phi\left[\log(m)\right]\ \dif m}{\int_{m=0.2}^{1.0} \Phi\left[\log(m)\right]\ \dif m} &&\label{eq:imfLog}\\[0.5em]
    && &= \frac{\int_{m=0.2}^{0.5} m\cdot X(m)\ \dif m}{\int_{m=0.2}^{1.0} m \cdot X(m)\ \dif m} &&\label{eq:imfLin}
\end{flalign}
where \(\Phi\left[\log(m)\right]\) in \cref{eq:imfLog} is the IMF in logarithmic mass units and \(X(m)\) in \cref{eq:imfLin} is the IMF in linear mass units.\par
\(\xi\) represents the ratio of low- to intermediate-mass stars. These mass ranges are expected to contribute to the observed spectrum of evolved galaxies, unlike higher-mass stars which are likely to have evolved into non-luminous remnants by the present-day. As presented in \cite{martin-navarro2019}, literature IMF formalisms from \cite{salpeter1955}, \cite{kroupa2002}, and \cite{chabrier2003} have \(\xi\) values of \(0.6370\), \(0.5194\), and \(0.4607\), respectively. Note here that these values are derived using the functional forms specific to each of these three IMF, which all differ from the function used in this work given in \cref{eq:imf}. These values merely provide a relative scale by which to compare with the \(\xi\) measurements we observe in our sample. Larger values of \(\xi\) represent greater relative contributions from low-mass stars, producing more dwarf-rich populations.\par
The greater computational efficiency of FIF is achieved by excluding regions of the spectrum outside of specific absorption features. This does in fact reduce its ability to constrain the variations in different regions of the spectrum, and the continuum itself may also contain useful information. To circumvent this issue, an initial fit is performed on the full spectrum using a (non-Bayesian) quadratic solver -- namely, {\tt pPXF} -- which provides a prior to the Bayesian fit for ages, metallicities, and abundances. Additionally, of the absorption features listed above, only \(\chemMgFe\) constrains the \(\chemAlphFe\) abundances. The variations of other individual elemental abundances are coupled to the variations of the \(\chemMgFe\), further improving the efficiency at the cost of reduced generality. Maps of stellar population properties for all passive ETGs in the \ftd\ sample derived using FIF are presented in \cite{martin-navarro2021a}. Here we use the derived maps of \(\xi\) from that work to build an orbit-based description of the IMF properties measured for our three-edge-on lenticular galaxies.

\subsection{An Orbital Analysis of the IMF}\label{ssec:orbIMF}
The method presented in \citetalias{poci2021} allowed the detailed analysis of a galaxy's assembly history through the intrinsic properties it provides. It exploits the straight-forward principle that the observed data result from the integrated contributions of many distinct populations of stars through the line-of-sight (LOS). We apply the same concept here, this time with the IMF parameter \(\xi\).\par
From dynamical models presented in \citetalias{poci2021}, the orbits were bundled into distinct dynamical sub-components using the orbital \(\lambda_z-R\) phase-space. While the same criteria for the phase-space bundling was applied to all three galaxies (see \citealt{poci2019}; \citetalias{poci2021}), the final set of sub-components depends on the specific phase-space distribution from the best-fitting \shw\ model of each individual galaxy. In the end, there are sets of \(448\), \(455\), and \(449\) distinct dynamical components with non-zero weights from the \shw\ models of FCC~153, 170, and 177, respectively. These numbers of phase-space components provide the required flexibility in physical space in order to reproduce the complex spatial variations in the projected stellar population maps. \par
The relative contribution (luminosity weight) from each sub-component/population \(i\) through the LOS in a given spatial bin is known via the decomposition of the orbital phase-space. There exists then a set of \(\xi^i\) which, when linearly combined with the pre-determined luminosity weights, reproduces the measured map of \(\xi\). This framework is described by \cref{eq:matrixIMF}.
\begin{equation}\label{eq:matrixIMF}
    \begin{pmatrix}
        \tilde{\omega}_1^1             & \tilde{\omega}_1^2             & \cdots & \tilde{\omega}_1^{N_{\rm comp.}} \\
        \tilde{\omega}_2^1             & \tilde{\omega}_2^2             & \cdots & \tilde{\omega}_2^{N_{\rm comp.}} \\
        \vdots                 & \vdots                 & \ddots & \vdots \\
        \tilde{\omega}_{N_{\rm aper}}^1 & \tilde{\omega}_{N_{\rm aper}}^2 & \cdots & \tilde{\omega}_{N_{\rm aper}}^{N_{\rm comp.}}
    \end{pmatrix} \cdot \begin{pmatrix}
        \xi^1 \\ \xi^2 \\ \vdots \\ \xi^{N_{\rm comp.}}
    \end{pmatrix} = \begin{pmatrix}
        \xi^\prime_1 \\
        \xi^\prime_2 \\
        \vdots \\
        \xi^\prime_{N_{\rm aper}}
    \end{pmatrix}
\end{equation}
The \(\tilde{\omega}\) represent the normalised luminosity weights for each dynamical sub-component, as defined in \cref{sec:data} (columns of the matrix), in each spatial aperture (rows of the matrix). The \(\xi^i\) represent the IMF parameter of each dynamical component \(i \in [1, N_{\rm comp.}]\), and the \(\xi^\prime_j\) represent the observed IMF parameter in each spatial aperture \(j \in [1, N_{\rm aper.}]\). The \(\xi^\prime_j\) are the observed values computed using the \(\Gamma_b\) measurements from FIF, as described in \cref{ssec:imfFIF}. We solve \cref{eq:matrixIMF} for the set of \(\xi^i\) by inversion. The outcome of this process is that each dynamical component has a fitted \(\xi\) which is constrained by the IMF parameter measured with FIF, and consistent with the orbital weights derived from the prior fit to the observed kinematics. This process works by exploiting the consistent weighting between the observations and the models; the observed stellar kinematics (and therefore the \shw\ model) as well as the outputs from the FIF technique are all luminosity-weighted. This means that the effect of a given component's intrinsic kinematics on the observed spectrum (the shifting and/or broadening of lines) is captured by the relative weighting of that same component's LOSVD, as derived by the dynamical decomposition of the \shw\ phase-space. Therefore, the intrinsic combination of sub-populations through the LOS is reflected in both the spectra for FIF and the dynamical decomposition. This same concept, but for age and metallicity, was verified in \citetalias{poci2021}.\par
Solving \cref{eq:matrixIMF}, which consists of a straight-forward non-negative linear matrix inversion, is done with a Bounded-Value Least Squares (BVLS) fit using the {\tt lsq\_linear} implementation within the {\sc SciPy} ecosystem \citep{virtanen2020}. Despite reducing the freedom of the model by grouping orbits in integral space into dynamical sub-components, there remains some level of degeneracy in the fit to the single map of mean \(\xi\). To minimise the impact of this, we apply linear regularisation to the solution weights, as described in detail in \cite{poci2019}. The regularisation, in the case where two solutions fit the data equally well, will favour the solution which has the smoothest distribution in weight-space; in our case, in the \(\lambda_z-R\) plane. In physical terms, orbits with similar angular momentum \((\lambda_z)\) and (cylindrical) radius will preferentially be given similar values of \(\xi\) {\em if} such a solution is otherwise indistinguishable from one which varies more rapidly.

\section{Results}\label{sec:resultsIMF}
\subsection{Fits to Observational Data}
\begin{figure}
    \centerline{\includegraphics[width=\columnwidth]{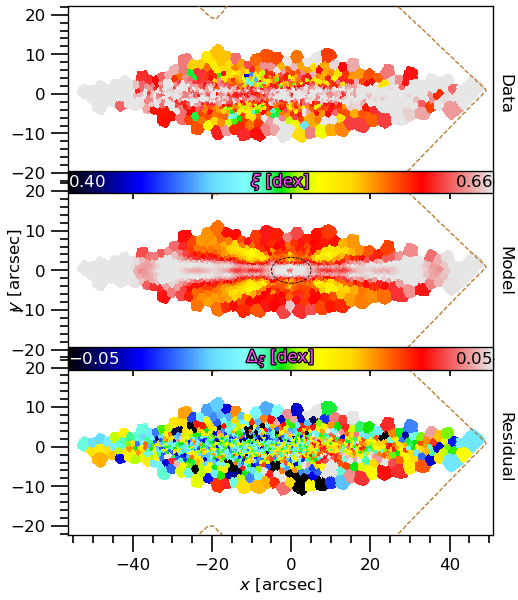}}
    \caption{The IMF fit for FCC~153. From top to bottom, the maps show the data, model, and residuals (data - model). The outline of the MUSE mosaic is shown in dashed brown. The dashed black ellipse in the `Model' panel illustrates the intrinsic radius which separates the `outer' and `inner' hot components as determined by the orbital phase-space (with ellipticity corresponding to the average ellipticity of the galaxy). Maps of \(\xi\) are presented on a common colour scale across all galaxies for comparison.}
    \label{img:mw153}
\end{figure}

\begin{figure}
    \centerline{\includegraphics[width=\columnwidth]{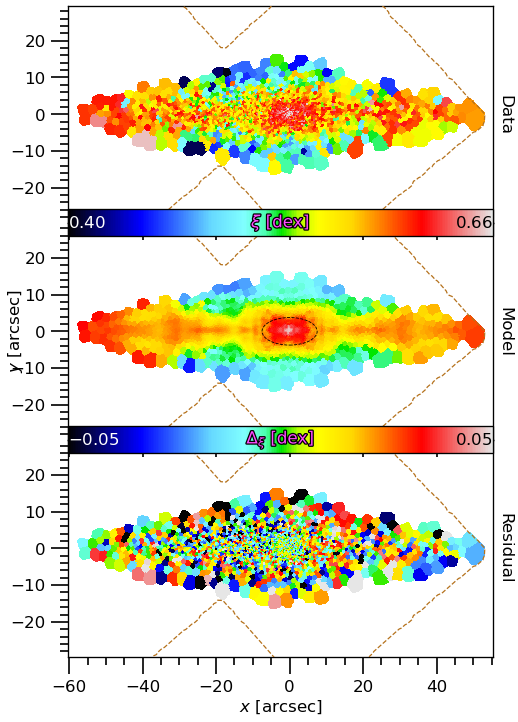}}
    \caption{As \protect\cref{img:mw153}, but for FCC~170.}
    \label{img:mw170}
\end{figure}

\begin{figure}
    \centerline{\includegraphics[width=\columnwidth]{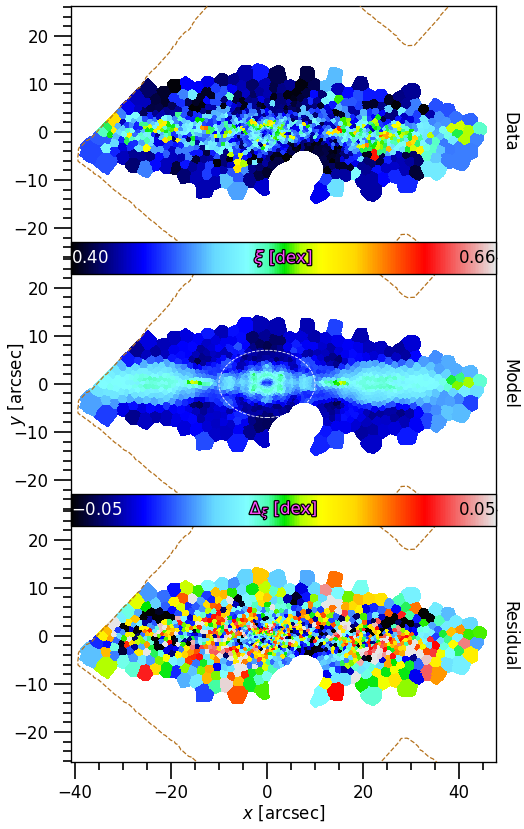}}
    \caption{As \protect\cref{img:mw153}, but for FCC~177.}
    \label{img:mw177}
\end{figure}

The results of the fits to the IMF parametrisations are presented in \cref{img:mw153,img:mw170,img:mw177}, for FCC~153, FCC~170, and FCC~177, respectively.
The structures in the \(\xi\) maps do not necessarily follow those of the age and metallicity maps directly \citepalias[luminosity-weighted measurements presented in][]{poci2021}. In FCC~153, \(\xi\) is noticeably elevated along the young, metal-rich disk. There is a minor central peak, but FCC~153 has a small (if any) central pressure-supported component, so a dramatic change in this region is not necessarily expected. In FCC~170, the IMF is seen to be dwarf-rich both in the relatively young disk and in the more metal-rich, old central component. Clearly, then, neither age nor metallicity alone can fully account for the IMF variation in FCC~170. Finally, in FCC~177, the IMF shows similar structure as the metallicity such that \(\xi\) is elevated along the disk and in a spheroidal-like central component. The exception is the very central region, embedded within the spheroid. This region appears to have a relatively high abundance of intermediate-mass stars (lower \(\xi\)). This feature is mirrored by significantly younger ages and lower \mlStar\ in this region, indicating a sudden shift in the time and conditions of that star-formation episode. However this region may be influenced by the young metal-rich nuclear star cluster in this galaxy \citep{fahrion2021}.

\subsection{The IMF in the Circularity Plane}\label{ssec:imfCirc}
\begin{figure}
    \centerline{\includegraphics[width=\columnwidth]{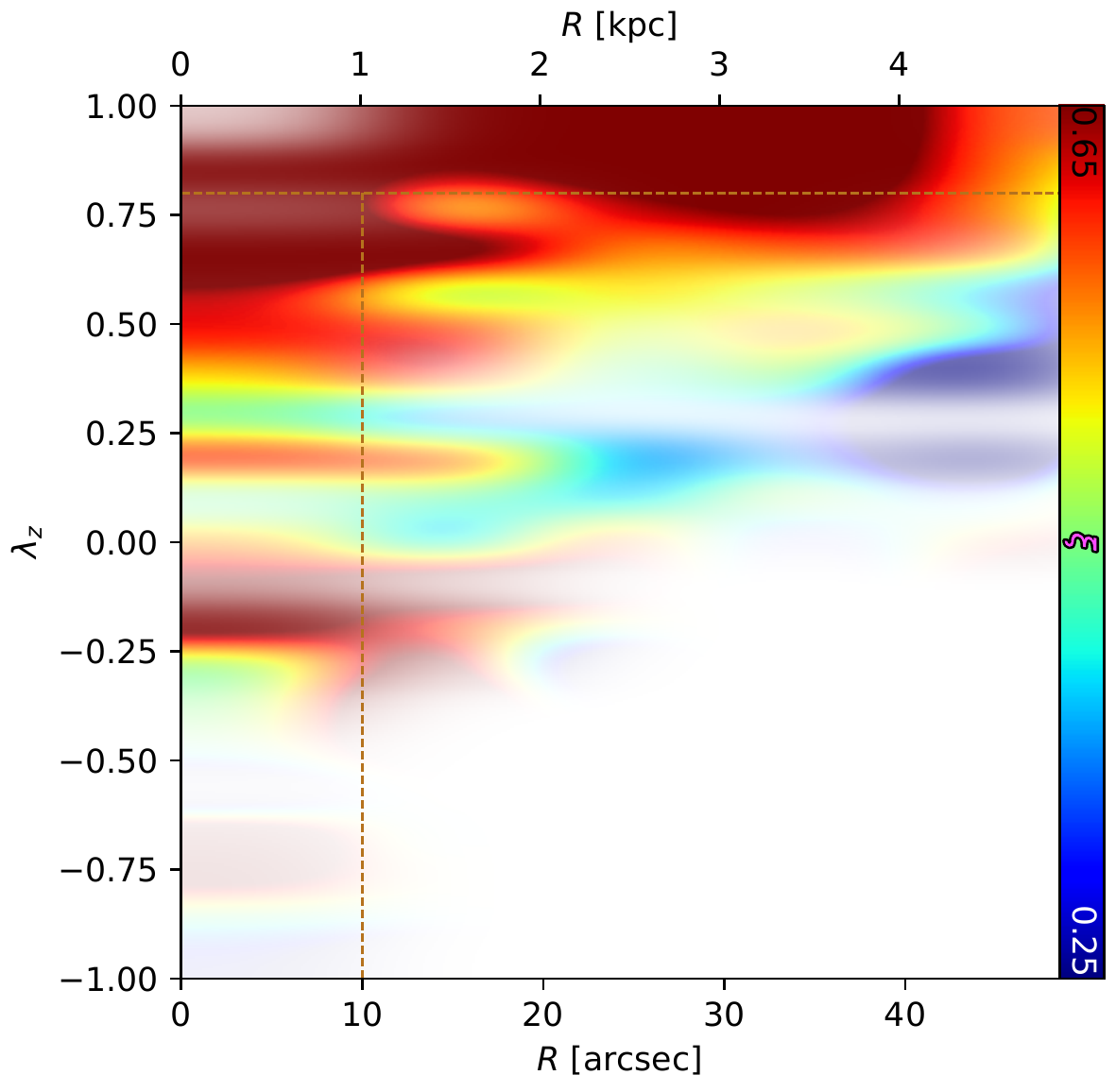}}
    \caption{The circularity phase-space coloured by IMF for FCC~153. The colour bar, common to all three galaxies, shows the luminosity-weighted average \(\xi\) in each region. The original weight distribution of the underlying orbits is depicted by transparency, where opaque represents high weight and transparent represents zero weight. The brown dashed lines illustrate the dynamical selection of the three broad components explored in \protect\cref{ssec:pgoc}.}
    \label{img:circIMF153}
\end{figure}

\begin{figure}
    \centerline{\includegraphics[width=\columnwidth]{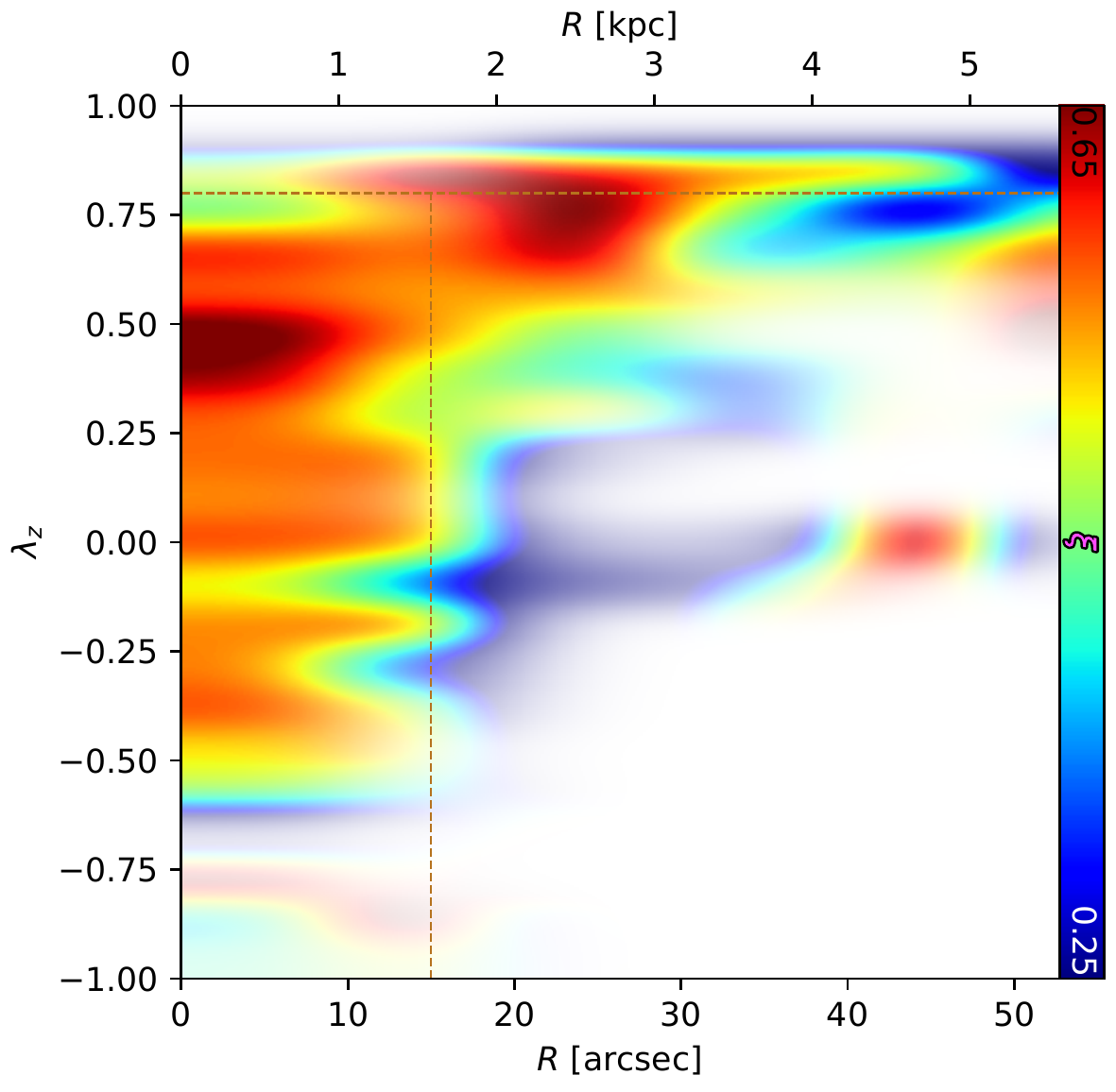}}
    \caption{As \protect\cref{img:circIMF153}, but for FCC~170.}
    \label{img:circIMF170}
\end{figure}

\begin{figure}
    \centerline{\includegraphics[width=\columnwidth]{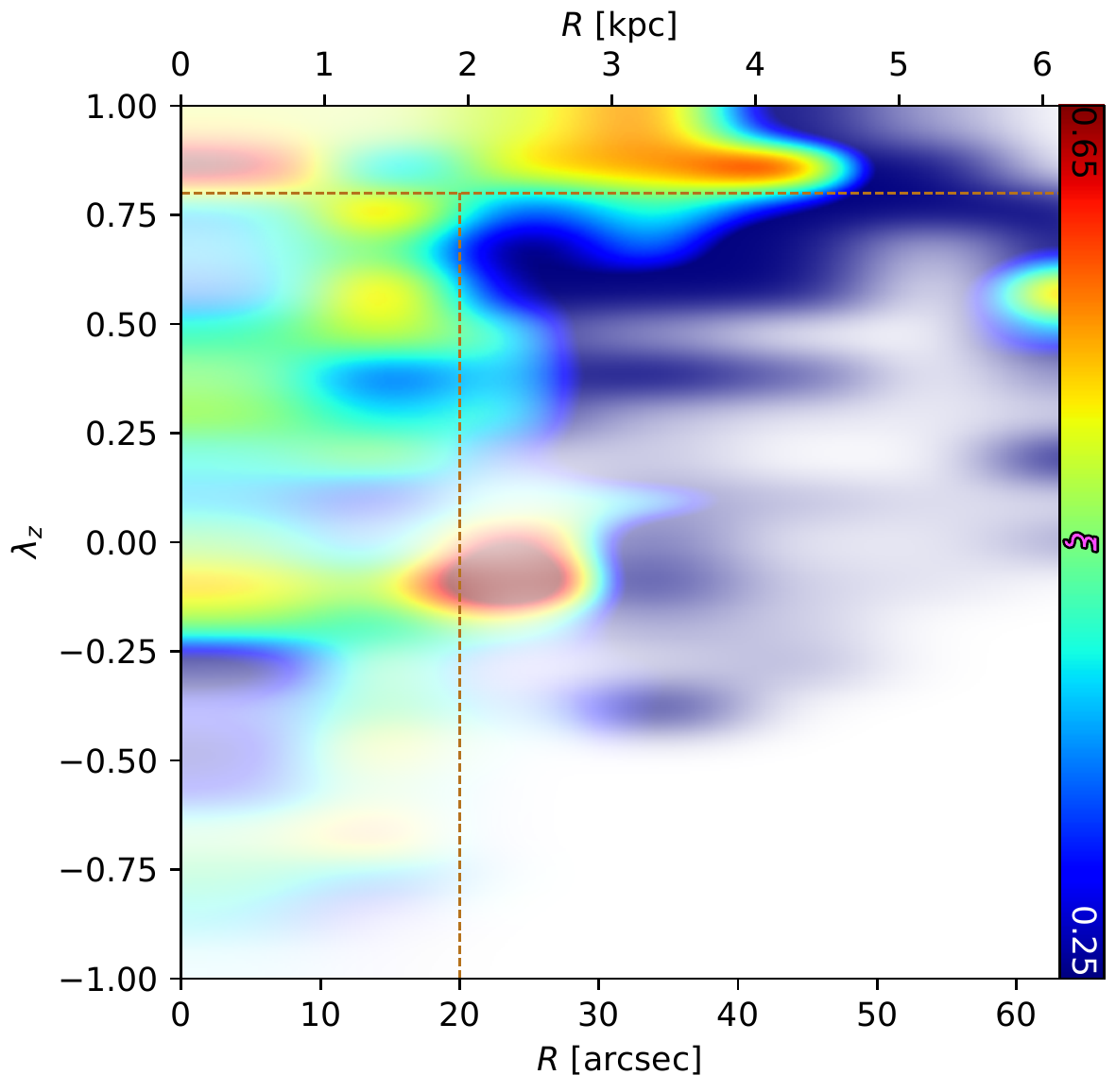}}
    \caption{As \protect\cref{img:circIMF153}, but for FCC~177.}
    \label{img:circIMF177}
\end{figure}

The orbital framework we have used to fit the measured \(\xi\) maps allows us to investigate the distribution of IMF throughout each galaxy; specifically, correlations with intrinsic dynamical properties. In \cref{img:circIMF153,img:circIMF170,img:circIMF177}, we re-project the circularity phase-space as a function of \(\xi\).
The transparency indicates the orbital weighting derived from the original \shw\ model. In these projections, regions with low transparency do not contribute significantly to the models/maps, irrespective of the integrated value of \(\xi\) (colour) in that region.\par
These projections clearly illustrate the relative difference of \(\xi\) between galaxies (most notably with FCC~177). Internally, they also show that the dwarf-rich (high \(\xi\)) populations exist on orbits with high angular momentum \((\lambda_z)\) and/or are centrally-concentrated. Conversely, the dynamically-warm regions always exhibit relatively less dwarf-rich populations within each galaxy. These differences may be related to the origin of the stars in each component, discussed further in \cref{sec:discuss}.

\subsection{The IMF of Principle Galactic Orbital Components}\label{ssec:pgoc}
Although many dynamical components were used to fit the observed \(\xi\) map, the dynamic range of the spatial variations in the case of the IMF are markedly reduced with respect to age and metallicity. As a result, we do not attempt a fine-grain dynamical decomposition of the IMF maps. Instead we draw conclusions about the major dynamical structures expected in \SZ\ galaxies in order to reduce the numerical noise of the fitting method caused by the relatively noisy observed IMF maps. To do this, we define a rotationally-supported `cold' component as having \(\lambda_z \geq 0.8\). A single radial cut is applied to the remaining orbits with \(\lambda_z < 0.8\) in order to isolate the central pressure-supported spheroid. This radius is derived from the original circularity phase-space by approximately identifying the natural ridge in the weight distributions given in \citetalias{poci2021} and illustrated by the transparency in \cref{img:circIMF153,img:circIMF170,img:circIMF177}. This is \(10\si{\arcsecond}\ (1.0\ \si{\kilo\parsec})\) for FCC~153, \(15\si{\arcsecond}\ (1.6\ \si{\kilo\parsec})\) for FCC~170, and \(20\si{\arcsecond}\ (1.9\ \si{\kilo\parsec})\) for FCC~177, reflecting the different dynamical configurations of the galaxies. These regions were previously identified as being bulge-dominated in \cite{pinna2019, pinna2019a}. We refer to these as the `inner' and `outer' hot components, respectively, in contrast to the `cold' component which occupies all radii at the highest circularities. These components are shown in \cref{img:circIMF153,img:circIMF170,img:circIMF177} for reference, and we can now investigate their respective average IMF properties.\par
\begin{figure}
    \centerline{\includegraphics[width=\columnwidth]{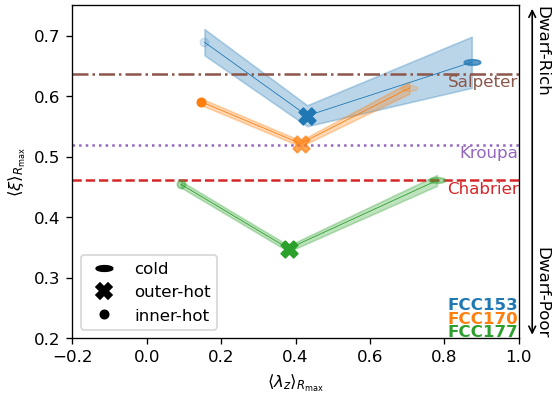}}
    \caption{The luminosity-weighted average IMF as a function of the luminosity-weighted average orbital circularity for the three dynamical components ({\em symbols}, defined in text) of the three \ftd\ galaxies. Transparency of the symbols denotes the relative orbtial weight of each dynamical component within a given galaxy, with opaque and transparent corresponding to high and low orbital weight, respectively. Horizontal lines mark literature IMF values for reference. Note that these reference values are computed using \cref{eq:imfLin} given their respective IMF functions, which differ from that used in this work, defined in \cref{eq:imf}. Only those orbits with time-averaged radii within the spectroscopic data are included during the averaging. Relatively dwarf-rich populations are present in the cold and inner hot components. The outer hot populations are markedly more dwarf-poor relative to the others. FCC~177 globally exhibits relatively dwarf-poor populations compared to the other two galaxies.}
    \label{img:imfcorr}
\end{figure}

To quantitatively compare the dynamical components and galaxies, we compute the luminosity-weighted average \(\xi\) for each region demarcated in \cref{img:circIMF153,img:circIMF170,img:circIMF177}. We show these averages in \cref{img:imfcorr} as a function of the average circularity of the dynamical components.
Specifically, for all orbits which satisfy the respective selection criteria of each dynamical component, we compute the average \(\xi\) weighting by the orbital luminosity weights from the original \shw\ model fit. Furthermore, to avoid being driven by orbits which are unconstrained by kinematics, all averages are computed for only those orbits within the spectroscopic FOV --- that is, with time-averaged radius within the maximum spectroscopic extent \(R_{\rm max}\).\par
Corroborating the conclusions from \cref{ssec:imfCirc}, \cref{img:imfcorr} shows that each outer hot component is deficient in dwarf stars relative to the other regions in its host galaxy. Moreover, \cref{img:imfcorr} clearly illustrates the absolute difference between FCC~177 and the other two galaxies. While our sample consists of three galaxies, the distribution of these galaxies in \cref{img:imfcorr} is inconsistent with being driven by many of their present-day properties. \(\xi\) does not correlate with either the present-day projected cluster-centric distance or suspected time of in-fall into the cluster, which both increase from FCC~170 to FCC~153. This is in agreement with the lack of dependence of the IMF on environment found previously, for both galaxy-scale \citep{eftekhari2019} and local \citep{damian2021} environment. There is also no correlation with present-day stellar mass or central velocity dispersion, both increasing from FCC~177 to FCC~170. Local \(\xi\) clearly does not correlate with the local orbital circularity, leading to the non-linear relation between the dynamical components of each galaxy in \cref{img:imfcorr}.\par
\begin{figure}
    \centerline{\includegraphics[width=\columnwidth]{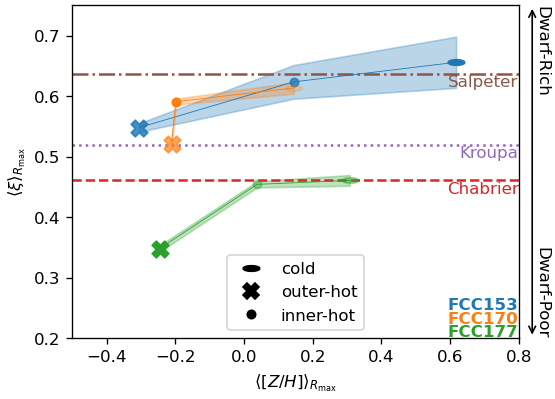}}
    \caption{As \protect\cref{img:imfcorr}, but for luminosity-weighted average metallicity.}
    \label{img:imfzet}
\end{figure}

We finally explicitly explore the relationship between local metallicity and local IMF in \cref{img:imfzet}.
There does indeed exist a mild correlation between the local metallicity and local IMF, as found previously \citep[e.g.][]{martin-navarro2021a}, such that more dwarf-rich populations are favoured for higher mean stellar metallicity (as a proxy for the ISM metallicity at the time of star formation). It can not, however, account for all of the IMF variations seen even in these few galactic components. For instance, the disk and inner hot component of FCC~177 exhibit the same average IMF, but different metallicities. Conversely, the inner and outer hot components of FCC~170 exhibit similar average metallicities but different IMFs. This is unsurprising, given the large scatter in the metallicity---IMF correlations, especially at low metallicity \citep[e.g.][]{martin-navarro2015, labarbera2019}.

\section{Discussion}\label{sec:discuss}
\subsection{The Local Variations of the Galactic IMF}
As seen from the results in \cref{sec:resultsIMF}, there appears to be no single galactic property which can account for the local variations of the IMF parameter. Instead, such variations are tied to the orbital structure of the host. This is unexpected from literature studies of IMF variations. There, the interpretation is usually that the `extreme' conditions present in the central regions of massive elliptical galaxies (often highly pressure-supported) is what gives rise to these IMF variations \citep[e.g.][]{sonnenfeld2012, wegner2012, dutton2013a, spiniello2014, martin-navarro2015, smith2015}. Moreover, there have only been a small number of studies in which the IMF is explored across galactic components. \cite{dutton2013} and \cite{brewer2014} use strong lensing and gas dynamics to constrain the IMF of bulges and disks separately. They find that the bulges favour dwarf-rich IMF such as \cite{salpeter1955}, and while the disk IMF are degenerate with the DM halo, they are more consistent with less dwarf-rich, Milky-Way-like IMF such as \cite{chabrier2003}.\par
We instead look to properties defining the conditions in which the stars actually formed, since present-day structural components do not necessarily capture evolutionary changes experienced by the galaxy. In \citetalias{poci2021}, one tentative conclusion was that FCC~170 and FCC~153 have accreted a similar amount of stellar mass, \(\logM[{\rm acc}] \sim 9\), while for FCC~177 it is approximately \(1\ \si{dex}\) lower. Specifically, for FCC~170 it is claimed to have been brought in by a higher number of lower-mass satellites in order to explain the lower metallicity of the warm orbital families, compared to fewer higher-mass satellites for FCC~153. These accreted populations would reside predominately in the outer hot components \citep[e.g.][]{monachesi2019, davison2021a}. If that is indeed the case, we may posit that the galaxy mass and/or interstellar medium (ISM) metallicity {\em at the time of star formation} could be the driver of these uncovered IMF trends, such that stars formed in the lower-mass/lower-metallicity progenitors contribute relatively dwarf-deficient populations to the present-day hosts. For FCC~177, being overall lower mass, this explains the global shift to less dwarf-rich populations in both the {\em in-situ} (cold and inner hot) and {\em ex-situ} (outer hot) components. Additionally, the offset of the outer hot component of FCC~170 towards less dwarf-rich populations with respect to that of FCC~153 in \cref{img:imfcorr} would be explained by the suspected lower-mass satellites accreted onto the former.\par
While investigating the role of turbulence in setting the slope of the IMF, \cite{nam2021} find that more shallow power spectra of turbulence result in more shallow high-mass IMF slopes. That is, they find that relatively lower power in large-scale turbulence correlates with relatively higher contributions from high-mass stars. \cite{chabrier2014} similarly find that in highly dense and turbulent environments, the peak of the IMF is shifted towards lower stellar masses, implying relatively higher contributions from low-mass stars. Should lower-mass galaxies have lower power in large-scale turbulence, this result would imply that they should have more shallow IMF slopes compared to higher-mass galaxies with relatively larger power in large-scale turbulence, supporting the results of our models. Interestingly, \cite{dutta2013} find only a weak relation between the dynamical mass (and \(\atomic{H}{i}\) mass) of spiral galaxies and the index of the power spectrum of turbulence, for dynamical masses between \(\sim 10^{11} - 10^{12}\ \si{\Msun}\). However, that sample covers a relatively small range of dynamical masses, so may not be conclusive. Potentially a more comprehensive sample would reveal any correlations between the dynamical mass of galaxies, the resulting turbulence of their ISM, and the subsequent impact on the dwarf-richness of the stars formed there.\par
Related, or perhaps alternatively, to turbulence, metallicity has been seen to clearly correlate with the IMF, both globally \citep[e.g.][]{martin-navarro2015} and locally \citep[e.g.][]{feldmeier-krause2021}. It is well-known that metallicity (both gaseous and stellar) correlates with mass \citep[e.g.][]{tremonti2004, gallazzi2005, gao2018a, curti2020}. However, recent evidence suggests that it is actually the average gravitational potential which drives these correlations rather than mass \citep[][using \(M_\star/R_\eff\) as an observational proxy to the gravitational potential]{barone2018, deugenio2018}, corroborated by correlations between average metallicity and central velocity dispersion \citep[e.g.][]{mcdermid2015}. The connection between the dwarf-richness of a population and the present-day gravitational potential of its host has already been proposed \citep{vandokkum2012}. However, this observational correlation would be `polluted' by the varying (and unknown) contributions from {\em ex-situ} populations in observed galaxies. An ideal test would require distinct accretion events to be identified and separated in both their IMF and metallicity distributions. Conversely, \cite{nipoti2020} find that dry minor mergers in massive ETG flatten existing gradients of \(\alpha_{\rm IMF}\), but they claim this is due predominantly to mixing of populations, rather than differences in the original populations of each progenitor galaxy.\par
The results of the literature works on massive elliptical galaxies are usually interpreted as being facilitated specifically by the conditions in the centres of galaxies, supporting the over-abundance of low-mass stars. However, by definition, these elliptical galaxies do not have any significant disk structure. Owing to their structural symmetry, the IMF is therefore studied radially, typical of gradients in other properties of ETG. Those galaxies also have the largest accretion fractions on average \citep{oser2010, khochfar2011, lackner2012, rodriguez-gomez2016}. Therefore, taken in the context of the results of this work, it is possible that the central regions of those massive ellipticals are dwarf-rich because they likely formed {\em in situ} in a relatively massive halo, while the exteriors are less dwarf-rich because their stars were formed in, then accreted from, lower-mass systems. This leads to a shift in the interpretation of local galactic IMF variations from being caused by the `extreme' conditions at the centres of massive elliptical galaxies, to being determined by the origins of the stars themselves, likely depending on the mass of the system in which they were formed. In present-day galaxies, IMF variations are, in this scenario, explained by the mixing of {\em in-situ} and {\em ex-situ} populations whereby the stars formed in a variety of host masses.\par
In absolute terms, the cold components measured here are significantly dwarf-rich, being consistent with a Salpter-like IMF for FCC~153 and FCC~170. This is comparable to the central regions of the massive ETG studied previously \citep[e.g.][]{sonnenfeld2012, wegner2012, dutton2013, shetty2014, oldham2018}, as well as the centres of a handful of other galaxies in the Fornax cluster \citep{martin-navarro2021a}, despite the significant physical differences in the galaxy types, densities, and internal kinematics. However, IMFs with `super-Salpeter' concentrations of dwarf stars have been seen in some of the most massive galaxies \citep[e.g.][]{tortora2013, spiniello2014, smith2015, conroy2017}, consistent with the idea that the gravitational potential in which the stars form is at least partially responsible for the level of dwarf-richness. Conversely, FCC~177 exhibits markedly less dwarf-rich populations globally, and especially in its outer hot region. FCC~177 has, however, had a significantly more extended and delayed SFH, and it is also the lowest mass, compared to the other two Fornax galaxies \citepalias{poci2021}.\par
In reality, the gravitational potential at birth is not likely the sole driver of observed IMF variations. There are, for instance, variations observed in galaxies expected to have experienced effectively no accretion \citep{ferre-mateu2017}. More likely, dwarf-rich populations could be favoured in a more metal-rich ISM at fixed gravitational potential (explaining the observed metallicity---IMF correlations and its large scatter), or deeper gravitational potentials at fixed ISM metallicity (explaining the deficiency of low-mass stars in the outer hot components in \cref{img:imfcorr}), at the time of star formation. But directly measuring such a correlation from observations would require both a detailed SFH in order to measure the historic mass and metallicity of the host galaxy during each episode of star formation, as well as timing every accretion event which has occurred in its assembly history.
\subsection{IMF Parametrisations}
There are a number of functional forms for the IMF presented in the literature. In addition to that of \cite{vazdekis1996}, other smooth functions have been proposed \citep[e.g.][]{miller1979, chabrier2003}, while many studies utilise generic power-laws with varying levels of flexibility in the slopes \citep[e.g.][]{kroupa2002, conroy2017, conroy2018, vandokkum2017, vaughan2018a, lonoce2021}. Moreover, for a given functional form, the IMF may be parametrised in any number of ways, with dwarf-to-giant ratios being frequently used \citep[e.g.][]{labarbera2013, lyubenova2016}. In this work, we have used a single parametrisation \((\xi)\) of a single functional form \citep{vazdekis1996}.\par
Although \(\Gamma_b\) is measured directly from the spectra via FIF, we do not use it in the orbital fitting. \(\Gamma_b\) describes the slope of an abstract mass distribution, and therefore does not satify the linearity requirements of \cref{eq:matrixIMF}. That is, for an integrated population through the LOS consisting of \(k\) sub-populations with \(\Gamma_b^{k}\), the slope of the integrated IMF is not given simply by the weighted average over all \(\Gamma_b^{k}\), because \(\Gamma_b\) responds non-linearly to changes in the mass function. Since \cref{eq:matrixIMF} is set up to compute the average through the LOS, \(\Gamma_b\) is unsuitable for this fitting process. Conversely, \(\xi\) represents a physical mass fraction, and thus the mass fraction of the total population is indeed given by the weighted average over its sub-populations. It is for this reason that \(\xi\) is used throughout this work.\par
For a given functional form of the IMF, \(\Gamma_b\) and \(\xi\) uniquely map to one another. However, different functional forms could produce the same \(\xi\) value with different characteristic slopes. This means that the precise translation from \(\xi\) to \(\Gamma_b\) (or indeed from \(\xi\) to specific spectral responses) is sensitive to the assumed form of the IMF. The main results of this work focus on the {\em relative} abundance of low-mass stars between dynamical components within galaxies for a given form of the IMF. However, comparisons to other results and literature IMF are subject to the assumption that the response of a spectrum for a given \(\xi\) value is the same for different IMF functions.
\subsection{Further Developments}
One implication of \cref{img:imfcorr} is that the greatest internal variations of IMF are in the `warm' components. These components reside in the diffuse outer regions, where there is limited coverage by the IFU data. Larger radial coverage of these galaxies will strengthen the results of this work, as the fraction of accreted material is expected to be higher at larger radius \citep[for instance,][]{karademir2019}. Further extending the IFU mosaic with additional (though necessarily deeper) observations, in particular along the minor axis, would also improve coverage of this important {\em ex-situ} material.\par
Broadly, our results indicate that both local and global galactic properties influence the IMF. In addition to those presented in \citetalias{poci2021}, we have found further independent evidence that the present-day orbits of stars retain information about their respective star-formation conditions, in this case encoded in their IMF. Larger samples of galaxies are certainly needed for more robust conclusions. If indeed the gravitational potential at the time of star formation is what determines the IMF, a targeted sample of so-called `relic' galaxies \citep{beasley2018} spanning a range of stellar mass to which we apply our methodology could in principle provide the necessary test of this hypothesis. This is because relic galaxies are suspected to have had little-to-no accretion to the present-day, which removes the uncertainty of measuring and timing the accretion events in a galaxy's assembly history. Incidentally, such a sample is currently being compiled \citep{spiniello2021}, however without sufficient spatial resolution to conduct the orbital analysis developed here. Independently, a sample in field environments mass-matched to the Fornax galaxies studied here would also be ideal to probe any potential effects from the cluster environment which may be present in the current results.\par
This implementation has enabled the direct comparison between the local orbital structure and stellar IMF for external galaxies. Our developments in fact pre-date the means to strictly verify them, as no current cosmological simulations can model the complexity of variable IMF -- though this is beginning to change with one recent instance which includes preliminary IMF treatments \citep{barber2018}. While isolated simulations have been performed incorporating non-universal IMF \citep{bekki2013}, testing our methodology on these simulations would only assess its accuracy in numerically recovering known input quantities -- which has already been shown in \citet{zhu2020};\citetalias{poci2021}. Thus, a more physically-motivated investigation of how accurately genuine IMF variations can be recovered within this orbital framework will require the advent of more sophisticated models for star-formation in future cosmological simulations.

\section{Conclusions}
This work has presented the first investigation of direct correlations between intrinsic local orbital properties and local stellar IMF. Using an orbit-based dynamical model of the stellar kinematics, we have reproduced the spatially-resolved observed maps of the stellar IMF. We then investigated how the stellar IMF depends on the intrinsic angular momentum \(\lambda_z\). We find that the relative abundance of dwarf stars, parametrised by \(\xi\), is higher in both the high-angular-momentum (`disk') and central pressure-supported (`bulge') orbits, while being markedly lower in the outer pressure-supported (`accreted inner halo') orbits. We interpret this relationship as being driven --- at least partially --- by the mass of the progenitor systems in which the stars formed, with lower-mass galaxies preferentially forming with lower relative abundance of low-mass stars. In this scenario, the variations of the IMF observed in our sample of external galaxies in the present-day is caused by the variations in individual assembly histories. This subsequently leads to a variety of {\em in-situ}-to-{\em ex-situ} population mixtures, where different populations were formed in progenitor systems of different mass and contribute different IMF to the galaxy observed in the present-day.\par
This analysis has presented an alternative interpretation of existing IMF results, which had inferred that significant IMF variations are only the result of extreme conditions in the most massive galaxies. We instead propose a scenario in which the IMF of a given population is determined by the level of turbulence supported by the gravitational potential of its host galaxy at the time of star formation, and that accretion/assembly processes can impose internal structure in the distribution of IMF properties within galaxies. Testing this hypothesis will require a dedicated sample of galaxies, likely facilitated by next-generation observational facilities, and/or highly-detailed simulations resolving the hydrodynamical effects on roughly parsec scales.


\section*{Acknowledgements}
Based on observations collected at the European Southern Observatory under ESO programme 296.B-5054(A). AP is supported by the Science and Technology Facilities Council through the Durham Astronomy Consolidated Grant 2020–2023 (ST/T000244/1). RMcD acknowledges financial support as a recipient of an Australian Research Council Future Fellowship (project number FT150100333). I. M-N and J. F-B acknowledge support through the RAVET project by the grant PID2019-107427GB-C32 from the Spanish Ministry of Science, Innovation and Universities (MCIU), and through the IAC project TRACES which is partially supported through the state budget and the regional budget of the Consejer{\'i}a de Econom{\'i}a, Industria, Comercio y Conocimiento of the Canary Islands Autonomous Community. GvdV acknowledges funding from the European Research Council (ERC) under the European Union’s Horizon 2020 research and innovation programme under grant agreement No 724857 (Consolidator Grant ArcheoDyn). EMC acknowledges support by Padua University grants DOR1885254/18, DOR1935272/19, and DOR2013080/20 and by MIUR grant PRIN 2017 20173ML3WW\_001. FP acknowledges support from grant PID2019-107427GB-C32 from The Spanish Ministry of Science and Innovation. Finally, we thank the anonymous referee for their detailed comments, which have improved the clarity of this work.\par
This work makes use of the \tfo{SciGar} compute cluster at ESO, and the \tfo{OzStar} supercomputer at Swinbourne University. The work also makes use of existing software packages for data analysis and presentation, including \tso{AstroPy} \citep{astropycollaboration2013}, \tso{Cython} \citep{behnel2011}, \tso{IPython} \citep{perez2007}, \tso{matplotlib} \citep{hunter2007}, \tso{NumPy} \citep{harris2020a}, the \tso{SciPy} ecosystem \citep{virtanen2020}, and \tso{statsmodels} \citep{seabold2010}.


\section*{Data availability}
No new data were generated or analysed in support of this research.



\bibliographystyle{mnras}
\bibliography{f3dII} 

\bsp	
\label{lastpage}
\end{document}